\documentstyle[aps,preprint]{revtex}
\newcommand{\be}{\begin{equation}}
\newcommand{\ee}{\end{equation}}
\newcommand{\bear}{\begin{eqnarray}}
\newcommand{\eear}{\end{eqnarray}}

\def\fun#1#2{\lower3.6pt\vbox{\baselineskip0pt\lineskip.9pt
 \ialign{$\mathsurround=0pt#1\hfil##\hfil$\crcr#2\crcr\sim\crcr}}}

\begin{document}
\draft
\title{KAON PHOTOPRODUCTION OF THE DEUTERON AND
THE $P$-MATRIX APPROACH TO THE $YN$ FINAL STATE
INTERACTION}
\author{B.O.Kerbikov} \address{ Institute of Theoretical and
Experimental Physics, 117218, Moscow, Russia} 

\maketitle

\vspace{2cm}
{\footnotesize
        Strangeness photoproduction of the deuteron is
        investigated  theoretically making use of the  covariant
        reaction formalism and the $P$--matrix approach the
        final state hyperon-nucleon interaction. Remarkably
        simple analytical expression for the amplitude is
        obtained. Pronounced effects due to final state
        interaction are predicted.}


\vspace{1cm}


\newpage

Up to now most of the  information on the hyperon--nucleon $(YN)$
interaction has been obtained either from hypernuclei or from
$K^-d$ and $\pi^+d$ reactions. After several decades of studies
our knowledge on the $YN$ system is still far from being complete.

Recently the interest to the
 $\Lambda N$ -- $\Sigma N$ system
has flared again in connection to the expected CEBAF
experimental results on the kaon photoproduction on the
deuterium [1]. The final state $YN(Y=\Lambda, \Sigma)$
interaction  (FSI) plays an
important role in the $\gamma d\to K^+YN$ reaction.
Therefore high resolution photoproduction experiments can
substantially deepen our understanding of  the $YN$ dynamics.

The problem of FSI in
 $\gamma d \to K^{+} Y n$ reaction has been addressed by several
 authors starting from the pioneering paper by
  Renard and Renard [2,3,4]. Two novel features differ the
 present work  from the previous studies.  First,
  covariant formalism based on direct evaluation of Feynman diagrams
  is used which allows to analyse the data beyond the region of low
  energy and low momentum transfer.
  Second,
  the $YN$ interaction is described within the
  $P$-matrix approach  which takes into
  account the subnuclear degrees of freedom and disentangle the
  dynamical singularities from kinematical threshold effects [5].
  The $P$--matrix analysis of the $YN$ interaction was presented in
  [6] (see also [7])
   and we shall use the set of parameters from [6].

    In our previous publication[8] we have presented some preliminary
   results without  proving the central assertion, namely that
  the FSI effects  allow remarkably simple evaluation within the
  $P$-matrix approach. The proof of this statement along with the
  presentation of the covariant reaction formalism contributes
  the core of the present publication.
  Tooled with the methods presented below one can easily treat
  FSI within any other approach making use of the known relation
  between $P$--matrix and potential approaches [9].

The double differential cross section of the reaction
$\gamma d \to K^{+} Y n$ reads
 \begin{equation}
  \frac{d^2 \sigma}{d\vert {\bf p}_{K} \vert d\Omega_{K}} =
\frac{1}{2^{11}\pi^{5}} \frac{{\bf p}_{K}^{2}}{k M_{d}E_{K}}
        \frac{\lambda^{1/2}(s_{2}, m_{Y}^{2}, m_{n}^{2})}{s_{2}} \int
        d\Omega^{*}_{Yn} \vert T\vert ^{2} \; .  \label{eq:dda-sigma}
           \end{equation}

Here $k$, ${\bf p}_{K}^{2}$, $E_{K}$ and $\Omega_{K}$ correspond to
the deuteron rest system  with $z$-axis defined by the incident
photon beam direction ${\bf k}$. The solid angle $\Omega^{*}_{Yn}$ is
defined in the $Yn$ center-of-momentum system, $s_2 =
(p_Y+p_N)^2,$
 $\lambda(x, y, z) = x^{2} - 2(y+z)x +
(y-z)^{2}$.

The fully covariant analogue of (1) valid in any  reference frame
has the form
 \be
  \frac{d^2
\sigma}{ds_2dt_1}=\frac{1}{2^{10}\pi^4\lambda(s,0,M^2_d)}
\int dt_2 ds_1
\frac{|T(t_1, s_1, s_2, t_2)|^2}{[-\Delta_4(t_1,s_1,s_2,t_2)]^{1/2}},
\ee
where $s=(k+p_d)^2,~
s_1=(p_K+p_Y)^2,~
s_2=(p_Y+p_N)^2,~
t_1=(k-p_K)^2,~
t_2=(p_d-p_n)^2,$
and $\Delta_4$ is a $4\times 4$ symmetric Gram determinat [10]. The
region of integration in (2) has to   satisfy $\Delta_4\leq 0$.
The number of essential final state Lorentz scalar variables is 4< namely
$t_1, s_1, s_2, t_2$.

The amplitude $T$ will be approximated by the  two  dominant
 diagrams, namely the tree (pole, or plane waves) graph and the loop
 (triangle) diagram with $YN$ FSI.Consider first the tree
 diagram.
  Two blocks  entering into it have to be
 specified:  {\em (i)} the elementary photoproduction amplitude
 $M^{\gamma K}$ on the proton, and {\em (ii)} the deuteron vertex
 $\Gamma_d$.  The elementary amplitude used in the present
calculation was derived from the tree level effective Lagrangian with
the account of several resonances in $s,t$ and chanels [11].
  This
amplitude has the following decomposition over invariant terms
[12]
\begin{equation}
 M^{\gamma K} = \overline{u}_{Y} \sum_{j=1}^{6}
 {\cal A}_j {\cal M}_{j}(s', t', u') u_{p} \; ,
 \label{eq:mgk}
 \end{equation}
  where $s'  = (k+p_{p})^{2} $, $t'  =
 (k-p_{K})^{2}$, $u' = (k-p_{Y})^{2}$.

 The decomposition of the deuteron  vertex function $\Gamma_{d}$ in
independent Lorentz structures has the form [13]

\begin{eqnarray}
\Gamma_{d}       & = & \sqrt{m_{N}}
\left[ (p_{p} + p_{n})^{2} -M_{d}^{2} \right]
     \left[\varphi_{1}(t_{2}) \frac{(p_{p}-p_{n})_{\mu}}{2m_{N}^{2}}
     + \varphi_{2}(t_{2}) \frac{1}{m_{N}} \gamma_{\mu} \right] {\cal
            E}^\mu(p_{d}, \lambda).
\end{eqnarray}

Here ${\cal E}^\mu(p_{d}, \mu)$ is the polarization
4-vector of the deuteron with momentum $p_{d}$ and polarization
$\lambda$.
 The vertex (4)  implies that the deuteron as well as the spectator
 neuteron are on mass shell while the proton  is off its mass shell.
 Now we can  write  the following expression for the tree diagram

\begin{equation}
	T^{(t)} = \overline{u}_{Y}
	\left\{ \left (\sum_{j=1}^{6} {\cal A}_{j} {\cal M}_{j} (s', t', u')
	\right )
	S(p_{p}) \Gamma_{d} \right\} u_{n}^c \; ,
	\label{eq:treeone}  
\end{equation}

\noindent where $S(p_p)$ is the proton  propagator and $u_{n}^c$ is a
charge conjugated neutron spinor.
 The deuteron  vertex in (5) may be substituted by the relativistic
 deuteron wave  function according to
 $\psi_d=[2(2\pi)^3 M_d]^{1/2} S(p_p)
\Gamma_d$ [14]. Then (5) can be rewritten as
\be
T^{(t)}= [(2\pi)^32M_d]^{1/2}M^{\gamma K}\psi_d,
\ee
where $\psi_d$ is the relativistic deuteron wave functions
discussed at length in [14], and
where summation over magnetic quantum numbers is tacitly assumed.

The loop  diagram  with $YN$ rescattering is given by the
expression
\begin{equation}
	T^{(l)} = \int \frac{d^{4} p_{n}}{(2\pi)^{4}} \;
	\overline{u}_{Y}(p'_{Y})
	\left\{ \left( \sum_{j=1}^{6} {\cal A}_{j} {\cal M}_{j} \right)
         S(p_{p})  \Gamma_{d} C S(p_{n}) T_{YN} S(p_{Y}) \right\}
         \overline{u}(p'_{N}) \; .
	\label{eq:tl1}  
\end{equation}

Here $C$ is the charge-conjugation matrix,
$T_{YN}$ is a  hyperon-nucleon vertex,
this vertex
being ``dressed'' by corresponding spinors constitutes the
hyperon-nucleon amplitude $F_{YN}$.

The comprehensive treatment of the loop diagram will be presented in
the forthcoming  publication while here we resort to a simple
approximation with the aim to exposure the effects of the FSI.
Only positive frequency components are kept in
the
propagators $S(p_n)$ and $S(p_Y)$ in (7), while the propagator $S(p_p)$
together with $\Gamma_d$ is lumped into the relativistic deuteron wave
function $\psi_d$ [14]. Then integration over the time component $dp^0_n$ is
performed. Thus we arrive at the following expression for $T^{(l)}$

\begin{equation}
T^{(l)} =  [{(2\pi)^{3} 2 M_{d}} ]^{1/2}
\int \frac{d{\bf p}_n}{(2\pi)^{3}}
\frac{M^{\gamma K} \psi_{d}(p_n) F_{YN}(E_{YN};q,q')}%
{{\bf q}^{2} - {\bf q}^{\prime 2} -i0} \; ,
	\label{eq:tl2}  
\end{equation}

\noindent where ${\bf q}$ and ${\bf q' }$ are the $YN$ c.m.
momenta  before and  after the     $YN$ FS, ${\bf q}={\bf
p}_n-\frac{1}{2} ({\bf k} -{\bf p}_k)$. The quality $F_{YN}(E_{YN};
qq')$ is the half-off-shell YN amplitude at $E_{YN}={\bf
q'}^2/2\mu_{YN}\neq {\bf q}^2/2\mu_{YN}$. The use of the
nonrelativistic propagator in (8) is legitimate since FSI is
essential at low relative YN momenta.  The arguments of the
elementary amplitude $M^{\gamma K}$ are specified in (3) but in the
kinematical region where $YN$ FSI is essential $M^{\gamma K} $ can be
considered as point-like and hence $M^{\gamma k}$ can be factored out
of the integral (8) at  the values of the arguments fixed by the
energies and momenta of the initial and final particles (i.e. at the
values of the arguments corresponding to the plane-waves diagram).

 Next we consider the $YN$ rescattering amplitude $F_{YN}$ and recall
 the interpretation of the  $P$-matrix in terms of the underlying
 coupled--channel quark--hadron potential [15].
 Namely, the $P$-matrix description is equivalent to the coupling
 between hadron and quark channels via the nonlocal energy dependent
 potential of the form [15]
 \be
 V_{hqh} =\sum_n\frac{f_n(r)f_n(r')}{E-E_n},
 \ee
 where $E_n$ are the energies of the six-quark "primitives" [9,5]
 ($P$-matrix poles), and the form-factors are given by
 $f_n(r)=\lambda_n\delta (r-b)$, where $b$ is the bag radius
 [9,5] and the coupling  constants $\lambda_n$ are related to the
 residues of the $P$-matrix via
 \be
 P=P_0+\sum_n\frac{\lambda^2_n}{E_n-E}.
 \ee
 As it was shown in [6]   a single primitive at $E_n=2.34 GeV$ is
 sufficient for a high quality description of the  existing $YN$
 experimental data.  In order to avoid lengthy equations we
 consider  in the  present note the region close to $\Lambda N$
 threshold while the preliminary results which included $\Lambda
 N- \sum N$ coupling were announced in [8] and will be treated
 within the present approach in a forthcoming detailed
 publication.

 The momentum--space half-of-shell amplitude $F_{YN}({\bf q},
 {\bf p}; E_{YN})$ corresponding to the potential (9) reads
 \be
 F( E_{YN}; q,  q')=-\lambda^2_n b^2\frac{\sin
 qb}{qb}d^{-1}(E_{YN})\frac{sin q'b}{q'b},
 \ee
 \be
 d(E_{YN})=E_{YN}-E_n+\frac{\lambda^2_n}{q} e^{iq'
b} \sin q'b.
 \ee
 This form of the amplitude
 allows to perform momentum integration
 in (8) analytically with an accuracy sufficient to display the
 effects due to FSI. In the complex $q$ plane the integral (8) picks
 up contributions from the poles of the propagator and the
 sigularities of the deuteron wave function. In a separate detailed
 publication we show both numerically  and using models for $\psi_d$
 that the contribution of the deuteron wave function singularities
 does not exceed 20\% (this conclusion can be immediately verified
 considering the simplest singularity of $\psi_d$ at
 $q^2=-\alpha^2)$. Thus substituting (11-12) into (8) and performing
 integration with the above remark in mind we get
 \be
 T^{(l)}\simeq [(2\pi)^32M_d]^{1/2} M^{\gamma K} \psi_d(p'_n)
 \left [\frac{1}{f(q')}-1\right ],
 \ee
 where $p'_n$ is related to $q'$ in the same way as $p_n$ to
 $q$ (see above), and where
  \be
 f(q')=1-\frac{\lambda^2_n b}{\Delta -q^{\prime 2}/2\mu_{YN}}
 e^{iq'b}\frac{ \sin q'b}{q'b},
 \ee
 and $\Delta = E_n-m_Y-m_N,~~ \mu_{YN}$ is the $YN$ reduced mass. One can
  easily verify that $f(q')$ is the Jost function corresponding to the
 potential (9) (some authors use the notation $f(-q')$). From (14) and (6) a
 remarkably simple expression for the sum of the tree and loop diagrams
 follows
 \be
 T^{(t)}+T^{(l)}=
 T^{(t)}/f(q'),
 \ee
 where the final state
 $YN$ momentum $q'$ is
 expressed in terms of
 the $YN$ invariant mass
 $s^{1/2}_2$ as
 $q'=\frac{\lambda^{1/2}
 (s_2, m^2_Y,
 m^2_n)}{2s^{1/2}_2}.$

  Expression (14) is
 physically  absolutely transparent: one immediately recognizes the standard
 enhancement factor [16] given by the inverse Jost function corresponding to
 the potential (9).  Inclusion of the $\Lambda N-\sum N$  coupling is
 straightforward [6]. The above equations may be used beyond the $P$-matrix
 approach to the $YN$ interaction since  there is a well trotted path
 connecting $P$-matrix and $S$-matrix approaches.

 In conclusion we present the results of the calculation obtained
 using equations
 covariant equation (2), Eqs. (6), and (14). Use  has been made of the
 elementary photoproduction amplitude from [11], the deuteron
 vertex function taken from the relativistic Gross model [14] and
 the plane-waves diagram  with this input was calculated in
 [17].  In Fig.1 the double differential cross section (1) is shown
 as a ~function~~ of ~~the ~~photon ~~energy~~ in ~the~ $\Lambda
 n$~~ invariant ~~mass ~~region ~close ~to ~the ~ threshold~~
 ($2.05 GeV \leq \sqrt{s_{\Lambda n}}\leq 2.10 GeV)$.  The
 enchanement of the cross section close to threshold due to
  FSI is quite pronounced. We remind that according to [6] the
  $^3S_1$ $\Lambda n$ scattering length is 1.84 fm in line with some
 versions of the Nijmegen potential [18].

 The author would like to thank  V.A.Karmanov for fruitful
 discussions  and suggestions. Valuable remarks by T.Mizutani and
 A.E.Kudryavtsev are gratefylly acknowledged. This work was
 possible due to stimulating contacts with C.Fayard, G.H.Lamot,
 F.Rouvier  and B.Saghai.  Hospitality and financial support from
 the University Claude Bernard and DAPNIA (saclay) as well as
 support from RFFI grant 970216406 are gratefully acknowledged.
 \newpage

 \begin{center}

{\bf \large REFERENCES}
 \end{center}

 \begin{enumerate}

\item
{\it Mecking B. et al.}// CEBAF Proposal PR-89-045, 1989.

\item
{\it Renard F.M.,   Renard Y.}// Nucl. Phys.  1967. V.B1. P.389.

\item
{\it Adelseck R.A.,  Wright L.E.}// Phys. Rev. 1989. V.C39 .
P.580;\\
{\it Li X., Wright L.E.}// J.Phys. 1991. V.G17. P.1127;\\
{\it  Cotanch S.R., Hsiao S.}// Nucl. Phys. 1986. V.A450. P.419 c;\\
{\it Lee T.-S.H. et al}// Nucl. Phys.
1998. V.A639. P. 247 c.
\item
{\it Yamamura H. et al.} Nucl.-th/9907029
V.2.

\item
{\it Bashinsky S.V., Jaffe R.L.}// Nucl. Phys. 1997. V. A625. P. 167.

  \item
   {\it Bakker B.L. et al.}// Sov. J. Nucl. Phys.  1986. V.43. P.982;\\
   {\it Kerbikov B. et al.}// Nucl. Phys. 1988. V.A480. P.585.

\item
{\it Heddle D.P., Kisslinger L.}// Phys. Rev.  1984. V.C30. P.965.

\item
{\it Kerbikov B.}//
  Strangeness photoproduction from the deuteron and
hyperon-nucleon interaction, hep-ph/9910361

\item
{\it Jaffe R.L., Low F.E.}//  Phys.  Rev.  1979. V.D19. P.2105.

 \item {\it E.Byckling and
K.Kajantie}// Particle Kinematics, John Willey and Sons, London-N.Y.,
1973.

\item
{\it David J.C. et al.}// Phys. Rev. 1996. V.C53. P.2613.

\item
{\it Chew G.F. et al.}// Phys. Rev.  1957. V. 106. P.1345.

  \item
   {\it  Gourdin  M. et al.}// Nuovo Cim.  1965. V.37. P.525.

  \item
   {\it Horstein J. and  Gross F.}// Phys. Rev.  1974. V.C10.
   P.1875;\\ {\it Buck W.W.,  Gross F.}// Phys. Rev.  1979. V.D20.
   P.2361.

 \item
{\it Simonov Yu.A.}// Phys. Lett. 1981. V. B107. P.1; Nucl.
Phys. 1987. V. A463. P.231.

  \item
  {\it Goldberger M.L. and Watson K.M.}// Collision Theory, John
  Wiley, N.Y., 1964.

\item
{\it Rouvier F.}//  Etude de la photoproduction d'etrangete sur le
deuton. These, Universite Claude Bernard, Lyon, 1997.

\item
{\it Nagels N.N., Rijken T.A. and de Swart J.J.}//  Ann. of Phys.
1973. V.79. P.338; Phys. Rev. 1977. V. D15. P. 2547; Phys.
Rev.  1979. V.D20. P.1633.
  \end{enumerate}

   \newpage

       Fig. 1. The double diffential cross section as a function of
         the photon energy for

         ~~~~~~~~$p_K=0.861 GeV$, $\theta_{\gamma K}=0^0$.  The dased
         line is the plane waves contribution,

         ~~~~~~~~  the
         solid line incorporates  FSI according to Eq. (14).

            \end{document}